\documentclass[pre,showpacs,floatfix,twocolumn]{revtex4}
\usepackage{graphicx}
\usepackage{amssymb}
\usepackage{natbib}
\usepackage{dcolumn}
\usepackage{bm}

\usepackage{color}

\usepackage{hyperref}
\usepackage{subfigure}

\newcommand{\der}{\ensuremath{\mathrm{d}}}

\usepackage{setspace}

\begin{document}

\title{ Tuning  spreading and avalanche-size exponents  in directed percolation with modified activation probabilities
}

\author{Fran\c cois Landes and Alberto Rosso}

\affiliation{CNRS-Laboratoire de Physique Th\'eorique et Mod\`eles Statistiques, 
Universit\'e Paris-Sud, 91405 Orsay, France}

\author{E. A. Jagla}

\affiliation{Centro At\'omico Bariloche and Instituto Balseiro, Comisi\'on Nacional de Energ\'{\i}a At\'omica, 
(8400) Bariloche, Argentina
}

\begin{abstract}

We consider the directed percolation process as a prototype of systems displaying a nonequilibrium phase transition into 
an absorbing state. The model is in a critical state when the activation probability is adjusted at some precise value $p_c$.
Criticality is lost as soon as the probability to activate sites at the first attempt, $p_1$, is changed. We show here that criticality can be restored by ``compensating" the change in $p_1$ by an appropriate change of the second time activation probability $p_2$ in the opposite direction. At compensation, we observe that the bulk exponents of the process coincide with those of the normal directed percolation process. However, the spreading exponents are changed, and take values that depend continuously on the pair $(p_1, p_2)$. We interpret this situation by acknowledging that the model with modified initial probabilities has an infinite number of absorbing states.
\end{abstract}

\maketitle

\section{Introduction}

There are many systems in nature that upon a continuous input of energy, react by sudden releases of the accumulated energy in the form of discrete events, that we call avalanches in general. Examples are the dynamics of sand piles, magnetic domain inversions in ferromagnets, stress release on the earth crust in the form of earthquakes, and many others. 
A remarkable characteristic of most of these realizations is the fact that the size distribution of the avalanches may 
display power laws that are a manifestation of the lack of intrinsic spatial scale in the system.

The theoretical analysis of such a variety of different processes has focused on the common features of the problems, and has tried to isolate the minimum necessary ingredients to explain the phenomenology that is common to most realizations.
There are numerous models that display critical behavior, and thus a power law avalanche size distribution. 
In most cases the obtained values of the exponents characterizing the avalanches are limited to a few possible values, corresponding to different universality classes. 

One of the reference models that are studied in this context is 
Directed Percolation (DP). DP is the paradigmatic example of dynamical phase transitions into absorbing states (see \cite{Henkel2008, Hinrichsen2006, Odor2004, Hinrichsen2000} for reviews).
It provides an example of a very robust universality class with well studied critical behavior, where power-law distributed avalanches are generated.
One of its most remarkable characteristics is its robustness:
many different particular models can  effectively be described within the DP scenario.

It has been shown that the critical properties of the DP transition are lost if the probability to activate a site for the first time is reduced with respect to the subsequent probabilities\cite{Rousseau1997, Jimenez-Dalmaroni2003}. 
In this paper we show that in this case, criticality can be restored by an appropriate increase of some of the following probabilities, in a process that we call ``compensation". Several critical exponents found at compensation do not coincide with those of pure DP. 
In particular, a time-reversal symmetry known to be valid for DP is violated at compensation. 
Other exponents conserve the values they have for DP. 
The values of the exponents that are seen to change depend in a continuous way on the 
precise choice of the activation probabilities.

The rest of the paper is organized as follows. In Sec. \ref{DP}, we review the critical properties of Directed Percolation, introducing the critical exponents and the scaling relations. In Sec. \ref{FIFA}, we recall the results known for a modified first infection model and present a variant: the modified first attempt model. In Sec. \ref{results} we present our results about the possibility of compensation. These results are discussed and compared with related models in Sec. \ref{discussion}. Finally our conclusions are in Sec. \ref{conclusion}.

\section{Critical Behavior of Directed Percolation}\label{DP}

DP is a dynamical model defined on a lattice, where  each site is associated with a state  (active or inactive) that evolves in time. 
Two commonly considered variants of this model are: site DP and bond DP.
In site DP, a site on the lattice will be active at time $t+1$ with probability $p$ if at least one of its neighbors is active at time $t$.
In bond DP, a site will be active at time $t+1$ with probability $1-(1-p)^k$, $k$ being the number of its active neighbors at time $t$. 
The configuration with no active sites is called an absorbing state because once it is reached, the dynamics stops.
In DP the absorbing state is unique.

For $p$ small, the system is trapped in the absorbing state exponentially fast, while for large $p$, the system has a finite probability to remain active indefinitely. 
There exists a threshold $p_c$ at which the system is critical, and in which the surviving probability decays to zero as a power law.
Around the threshold $p_c$ the system displays a non equilibrium phase transition from a fluctuating phase to an absorbing state.
As for standard equilibrium phase transitions,  universal behavior and critical exponents are expected.
It was found that both site and bond DP belong to the same universality class.
Here we focus on bond DP on a two dimensional square lattice, for which $p_c \simeq 0.287338$ \cite{Dickman1999}.

As $p$ is the control parameter of the transition, we denote the distance from criticality as $\Delta \equiv |p-p_c|$.  Two different order parameters can be defined:
When  the initial condition corresponds to a fully active lattice the relevant question is to determine the density of active sites when $t \to \infty$ (the stationary state), namely $\rho_{\text{st}}$. For $p<p_c$,  $\rho_{\text{st}} =0$, for $p>p_c$, $\rho_{\text{st}} =\Delta^\beta$. When at time $t=0$ a single site located at the origin is active, a cluster of active sites spreads from it. Here the relevant question is to determine  the probability to remain out of the absorbing state  when $t \to \infty$, namely $Q_{\text{st}}$. For $p<p_c$,  $Q_{\text{st}} =0$, for $p>p_c$, $Q_{\text{st}} =\Delta^{\beta^\prime}$.

Similarly to the case of equilibrium phase transitions, when approaching criticality,  a diverging length $\xi_\perp \sim \Delta^ {-\nu_\perp}$ describes the spatial correlations. Moreover in dynamical phase transitions there is a characteristic scale for time correlations, $\xi_\parallel \sim \Delta^ {-\nu_\parallel}$. These scales are independent of the observable and thus of the initial condition, while one expects the two distinct order parameters $\rho_{\text{st}}$ and $Q_{\text{st}}$ to be characterized by different exponents $\beta$ and $\beta^ \prime$
\cite{Footnote1}. We will see that other quantities display power law behavior with different critical exponents, however it is possible to write scaling relations that constrain the set of critical exponents to only four  independent quantities.

In practice, in numerical simulations it  is convenient to start from the single seed initial condition and let the cluster evolve up to time $t$. 
To characterize the growth of spreading clusters, one measures the survival probability $Q(t)$ and the average number of active sites at time $t$, $N(t)$. These two quantities obey the scaling forms:
\begin{eqnarray}
Q(t) & \propto &t^{-\delta}g_1(t/\xi_\parallel)  \label{Q1}\\
N(t)& \propto &t^{\eta}g_2(t/\xi_\parallel)
\end{eqnarray}
where $g_1$ and $g_2$ are $1$ at  $t=0$, and 
$g_i(x)\to 0$ for $x \to \infty$ 
below threshold. 
When we consider surviving clusters only, we can measure the average spatial extension of the cluster at time $t$, namely $L^d(t)$, and the average density $\rho(t)$ of active sites at time $t$ inside this region. These two quantities obey the scaling forms:
\begin{eqnarray}
\rho(t) & \propto &t^{-\theta}g_3(t/\xi_\parallel)  \\
L(t)& \propto &t^{\frac{1}{z}}g_4(t/\xi_\parallel)
\end{eqnarray}
where $g_3$ and $g_4$ behave similarly to  $g_1$ and $g_2$ below threshold.

Above threshold, both $Q(t)$ and $\rho(t)$  approach their asymptotic stationary state, $Q_{\text{st}}$ and  $\rho_{\text{st}}$, at a characteristic time $\sim \xi_\parallel$, so that two scaling relations can be written:
\begin{eqnarray}
\beta= \theta \nu_\parallel  \\
\beta'= \delta \nu_\parallel 
\end{eqnarray}
At the  critical point the scale invariance predicts that if time is rescaled by a factor $b$, space should be rescaled by a factor 
$b^{\nu_\perp / \nu_\parallel}$. Thus the size of a cluster grows as $L(t) \sim t ^ {\nu_\perp / \nu_\parallel}$ and a third scaling relation can be written:
\begin{equation}
 z=\frac{\nu_\parallel }{ \nu_\perp}
\end{equation}

Finally a \textit{generalized hyperscaling relation}\cite{Mendes1994a} valid below the upper critical dimension \cite{Hinrichsen2000} relates the four quantities previously defined.
Namely $N(t)$ can be expressed as the the sum of two contributions: the active sites of surviving clusters ($\sim \rho(t) L^d(t)$) which have probability $Q(t)$, and the contribution of dead clusters.
 This writes as:
\begin{eqnarray}
N(t) &=& L^d(t) \rho(t) \cdot Q(t) + 0 \cdot (1-Q(t))   \nonumber\\
\eta &=& \frac{d}{z} -\theta -\delta. \label{hypers}
\end{eqnarray}

Below threshold, each cluster can be identified with an avalanche and dies in a finite time $T$.  We define the size $S$ of an avalanche as the total number of activations that occurred, and are mainly interested in its statistics,
 $P(S)$, which is expected to follow a power law at criticality: $P(S) \sim S^{-\tau}$. 
The characteristic size of an avalanche is related to $T$ through 
\begin{equation}
\label{LL}
S(T) \sim \int_0^T \frac{N(t)}{Q(t)} \der t \sim T^{1+ \eta + \delta }
\end{equation}
Assuming that fluctuations around this characteristic value are small, we can write $P(S) \, \der S \sim - Q^\prime(T)  \,\der T$
where $- Q^\prime(T) \sim T^{-\delta-1}$  stands for the rate of death. Combining the latter relation with Eq.(\ref{LL}) we have
$P(S) \sim T^{-(1+\eta+2\delta)}
\sim S^{- \left( \frac{1+\eta+2\delta}{1+\eta+\delta}  \right) }$, and a scaling relation for the exponent $\tau$ can thus be written:
\begin{equation}
 \tau = \frac{1+\eta+2\delta}{1+\eta+\delta} = 1 + \frac{\delta}{1+\eta+\delta}. \label{tau}
\end{equation}

\begin{figure}
\includegraphics[width=0.5\textwidth]{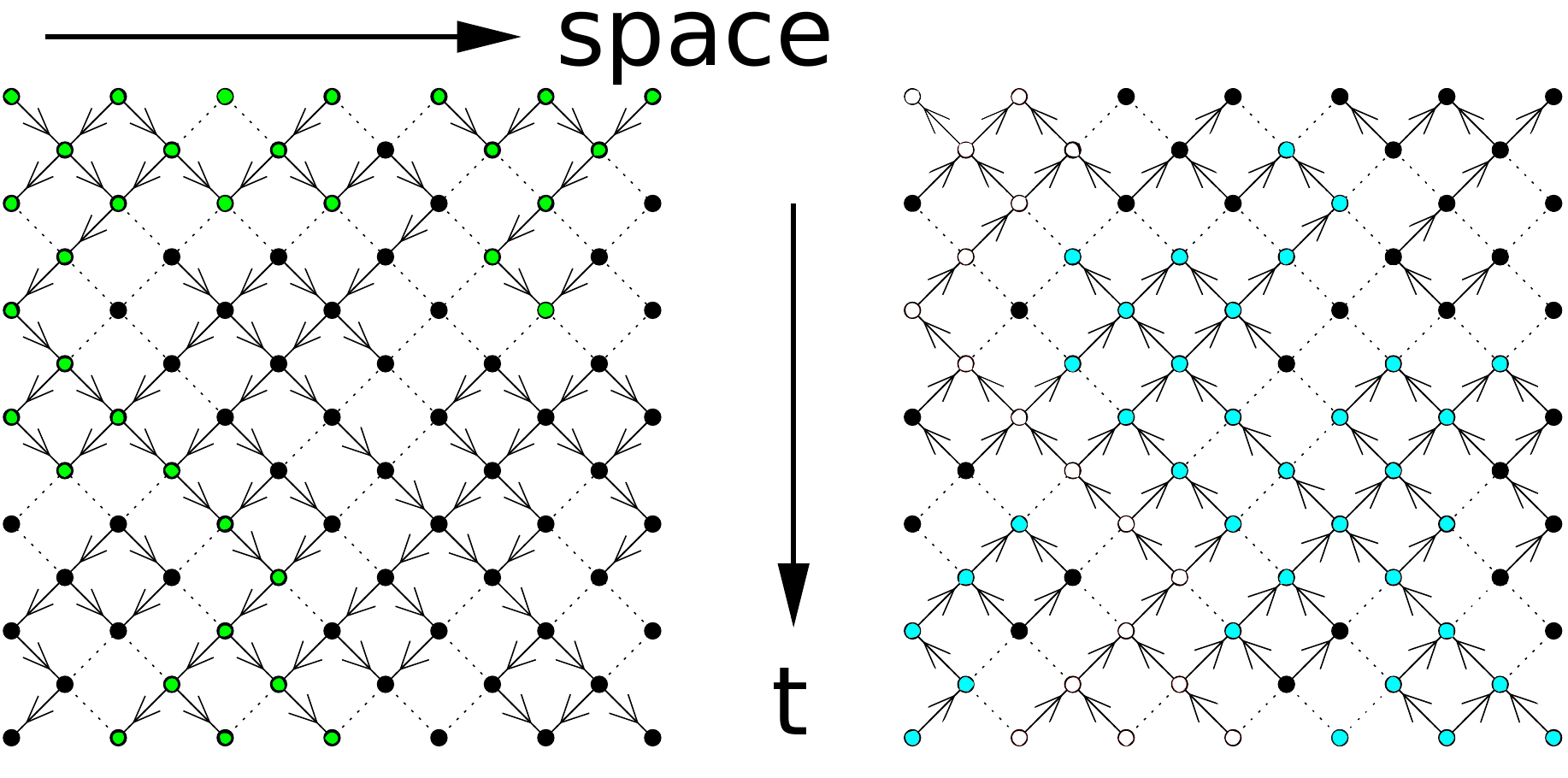}
\caption{(Color online). $1$-dimensional bond DP. Normal direction of time is downwards. The arrows are given once and for all and are the same for both panels. Final time is $t=12$.
Left: DP starting with a fully active lattice: $M=7$ occupied green (light gray) seeds, with final density $\rho(t) = \frac{3}{7}$. Right: DP with time reversed. In light blue (light gray), the paths which die before the end. In open circles (white), the paths that survive until $t$. There are exactly $m=3$ seeds that participate in surviving walks: $Q(t) = \frac{3}{7}$.
}
\label{timeR1}
\end{figure}

The exponents and relations that we introduced here are general features of all absorbing phase transitions, which are characterized by only four independent exponents: $\delta, \theta, z$ and $\nu_\parallel$. 
The exponents $\beta, \theta, \nu_\parallel,\nu_\perp $ and $z$ are called ``bulk exponents" because they can be measured both from the fully active initial condition, and from the single seed initial condition with averages performed over surviving runs exclusively.
The exponents $\beta^{\prime}, \delta, \eta$ and $\tau$  are called ``spreading exponents" because they are measured  starting from a single seed, with averages performed over all runs.

DP has an additional symmetry associated with time reversal, which implies that $\theta=\delta$ \cite{Grassberger1979, Hinrichsen2000}.
This is schematically proved in Fig. \ref{timeR1} for $1$-dimensional bond DP where arrows, drawn with probability $p$, connect neighboring sites.
An activated site at the start of an arrow activates the site at the end of the arrow.
The key observation is that the direction of time is arbitrary: starting from the top 
is equivalent to starting from the bottom with reversed arrows. The survival probability $Q(t)$ with fully active initial condition (with normal direction of time) is exactly equal to the density $\rho(t)$ with single seed initial condition in reversed time.
This exact relation thus reads:
\begin{eqnarray}
Q(t) &=&\rho(t) \\
\delta &=& \theta.
\end{eqnarray}
This is exact for bond DP, while in general $Q(t) \sim \rho(t)$, thanks to the universality of DP.
A necessary condition for this time-reversal symmetry is the uniqueness of the absorbing state. In a process with multiple absorbing states, or ageing, one cannot freely reverse the arrow of time.

We recall $2$-dimensional DP exponents precisely measured in numerical simulations  \cite{Dickman1999}:
\begin{eqnarray}
\label{pureDP}
\delta=\theta= 0.4505 \pm 0.001  &\;& z=1.766 \pm 0.002  \nonumber\\
\nu_\parallel=   1.295\pm 0.006      &\;&   \eta= 0.2295 \pm 0.001.
\end{eqnarray}

\section{First Infection and First Attempt Models}\label{FIFA}
\begin{figure}
\includegraphics[width=0.5\textwidth]{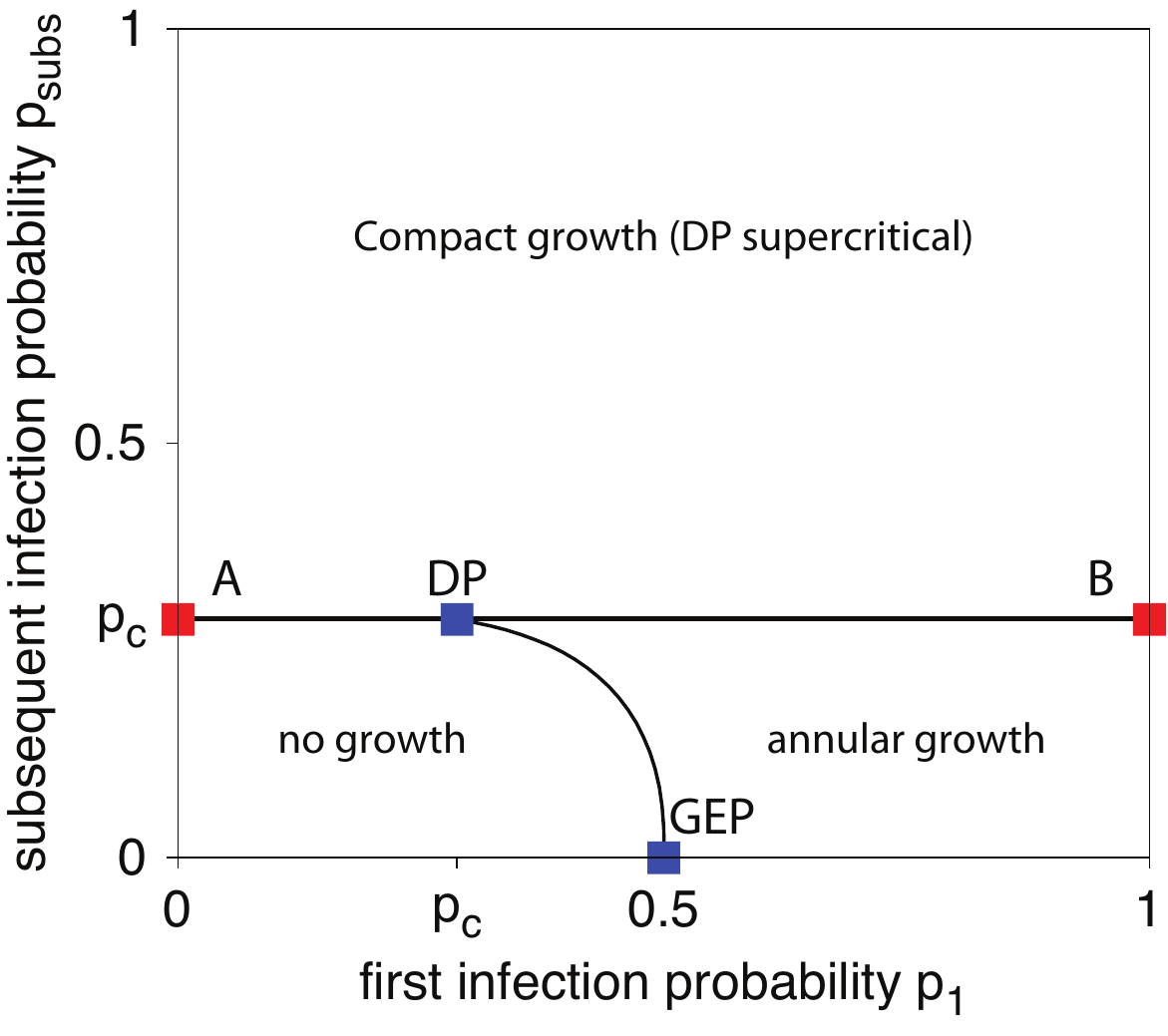}
\caption{(Color online). Phase diagram of the model with a first infection probability $p_1$ different from the   subsequent reinfection probability $p_{\text{subs}}$ .
}
\label{phased}
\end{figure}

A generalization of the bond DP process is the modified first Infection Model (IM) \cite{Rousseau1997,Jimenez-Dalmaroni2003, VanWijland2002,Dammer2003}. In this variant,
the probability to activate a site for the first time is given a value $p_1$ different from the value of the subsequent activations (that we call $p_{\text{subs}}$). 
This  has been considered as a model to describe epidemic processes with partial immunization.
In this context, the activation of a site is called {\em infection}, and it is understood that the possibility of the subsequent reinfection probability $p_{\text{subs}}$ can differ from the first infection probability $p_1$ due to ``immunization" effects. 
The phase diagram of this problem in $d=2$ was established in \cite{Rousseau1997} and is reproduced in Fig. \ref{phased}, for reference.

DP critical behavior occurs at $p_1=p_{\text{subs}}=p_c$. At this point, 
$Q(t)$, $N(t)$, $L^2(t)$ and $\rho(t)$ have the power law distribution corresponding to pure DP.
In the line terminated in the blue (darkest) points, the system experiences a phase transition corresponding to the
so called General Epidemic Process (GEP). The fixed point of (bond) GEP is located at $p_1=1/2$, $p_{\text{subs}}=0$ and corresponds exactly to the problem of bond Isotropic Percolation.

Along the AB line, except for the unstable DP fixed point, the system is not critical. 
In particular, the surviving probability $Q(t)$ 
and the size distribution of the avalanches $P(S)$ 
decays faster than a power law.
The instability of the DP fixed point was shown in \cite{Jimenez-Dalmaroni2003}: the renormalization flow takes one from any point in AB (outside the DP point) to either A or B.

Instead of the case in which there are different probabilities for the first infections, 
we will focus in this paper on the case in which different probabilities occur for
successive {\em attempts}, namely irrespective if the activation of the site actually occurred or not. 
The state is defined by the number of trials of activation, not the number of infections.
The reason to study this variant is that it may be useful to understand the avalanche size distribution in some models of seismic phenomena (the connection of DP with this problem will be discussed elsewhere).
We will refer to this variant as the Attempt Model (AM), to distinguish it from the 
Infection Model (IM) previously described.
The AM is a sort of milder modification of the original DP problem, compared to the IM. 
We expect the phase diagram of the AM to be qualitatively similar to that of the IM.

In particular, the DP fixed point is clearly located at the same position,
while the GEP point is slightly different. As we stated before, for the IM the GEP point corresponds to $2$-dimensional \textit{bond} Isotropic Percolation ($p_1=0.5$, $p_{\text{subs}}=0$). Instead, for the AM it corresponds to $2$-dimensional \textit{site} Isotropic Percolation ($p_1\simeq 0.592746$, $p_{\text{subs}}=0$).
Indeed, we observe that for the AM, when $p_{\text{subs}}=0$, a site can be activated only at the very first attempt, with probability $p_1$ (no matter if we consider site or bond DP), 
thus the sites that are activated once with this rule are exactly the sites activated in $d$-dimensional site Isotropic Percolation.



The main difference between AM and IM is that the AM has a non-singular limiting behavior as $p_1\to 0$, leading in particular to a finite mean event size $\langle S \rangle$ in this limit, whereas 
for the IM $\langle S \rangle$ goes to $0$ as $p_1\to 0$. 

\section{Recovering Criticality with Compensation: Model and Results}\label{results}

The main point addressed in the present paper is to show that for the AM the lack of criticality generated by a value of the first attempt $p_1$ smaller (larger) than $p_c$ 
can be  ``compensated"
by a larger (smaller) than $p_c$  
second attempt probability $p_2$. 
We will present strong numerical evidence showing how this compensation occurs, restoring critical behavior in the system. 

In addition, a remarkable result is that at compensation, several critical exponents of the problem, 
in particular the bulk exponents $\theta, z$ and $\nu_\parallel$, take their normal DP values, 
while the spreading exponents ($\delta, \eta, \tau$) depend on the precise values of $p_1$ and $p_2$. 

We will not discuss the possibility of compensation in the IM since we cannot be conclusive at present. 
Although it seems that compensation can be obtained, numerical evidence is not enough for a discussion on the variation or not of the obtained critical exponents.
We prefer to concentrate on the Attempt Model, where we are much more confident with the numerical results.

We consider the case in which the first two \textit{attempts} $p_1$ and $p_2$ differ from the subsequent ones, 
that from now on we consider to be equal to the critical DP value: $p_{i>2} \equiv p_{\text{subs}}=p_c=0.287338$. 

A heuristic argument suggesting that such a compensation can result in criticality is the following.
As a perturbation, the relevant character of a change in  $p_1$ was demonstrated in \cite{Jimenez-Dalmaroni2003} for the IM.
The analysis presented there indicates that a change in $p_2$ generates qualitatively the same kind of perturbation 
(to leading order) 
than a different $p_1$. 
Therefore, it is not surprising that there are particular combinations of $p_1$ and $p_2$ at which the leading term of both perturbations cancel each other. 
These particular combinations will be the compensating pairs of values $(p_1,p_2)$. However, the fact that we do not recover the pure DP exponents indicates that higher order terms do not vanish, but result in a marginal perturbation. 

We present first the numerical evidence of the compensation effect. 
In all simulations, 
we started from a single active site (seed) a time $t=0$ that was in a state of being attempted twice, and let the clusters grow until their natural death, or time $t=10^5$ or $10^6$. The lattice is large enough so that the boundaries are never reached by the cluster.
To be very precise about our choices: a site that has been successfully infected at the first attempt is still in a state of being attempted just once.

\begin{figure}
\includegraphics[width=0.5\textwidth ]{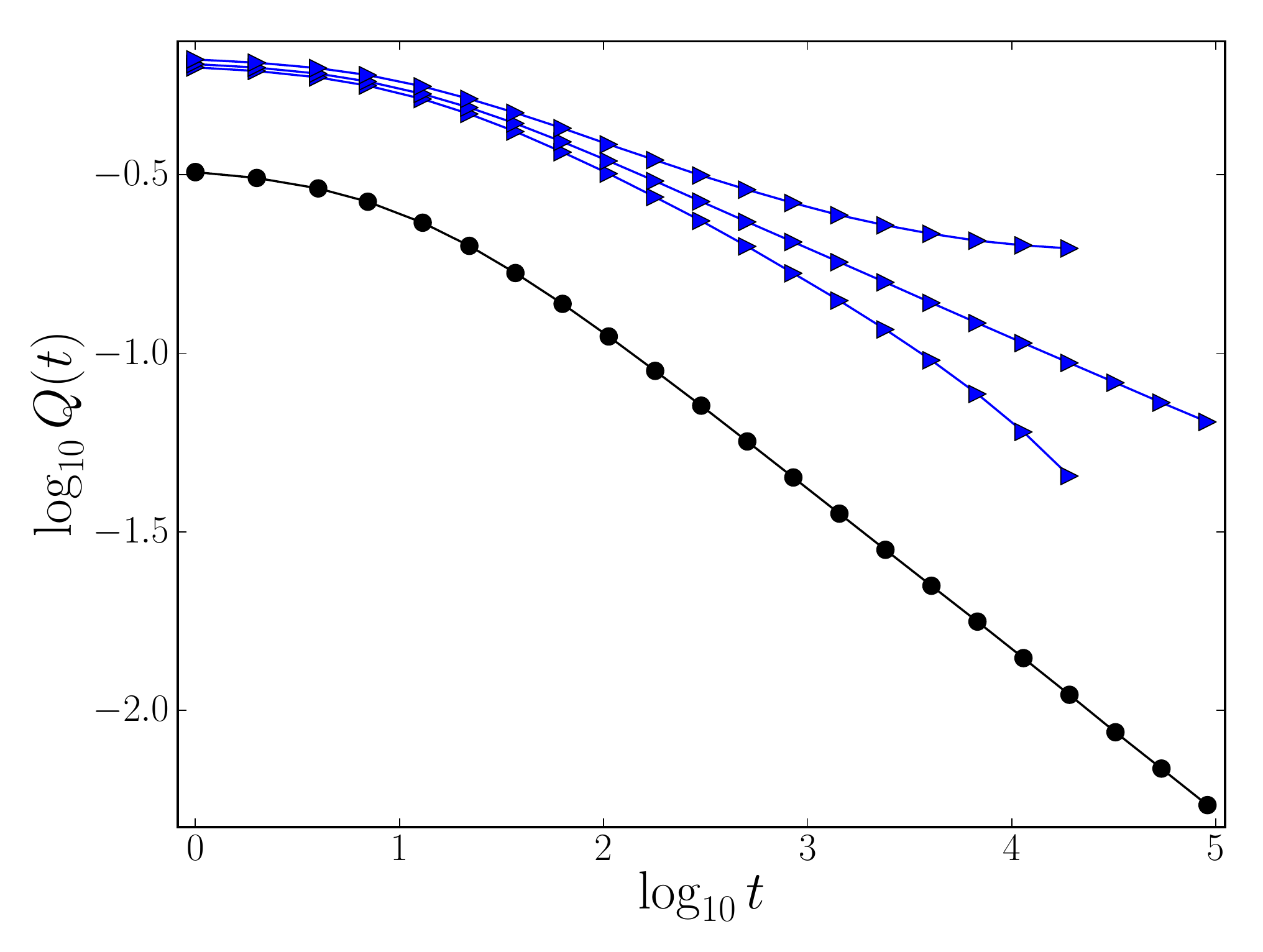}
\caption{(Color online). 
$Q(t)$ for different choices of $(p_1,p_2)$. 
Circles represent the pure DP at $p_1=p_2=p_{\text{subs}}=p_c$. Averages are performed over $10^6$ samples.
Triangles represent the AM with $p_2=0, p_{\text{subs}}=p_c$. 
From top to bottom, we used $p_1 = 0.494, 0.4888, 0.485$. For $p_1=0.4888$, $Q(t)$ displays a clear power-law with $\delta=0.25 \pm 0.01$. Averages were performed over about $10^5$ samples.
}
\label{qt1}
\end{figure}

We investigated two pairs of compensating points $(p_1,p_2)$ and compared with usual DP (in which $p_1=p_2=p_{\text{subs}}=p_c$).  
For the first one, we set $p_2=0$ and varied $p_1$ in order to find the critical point. 
In Fig. \ref{qt1}, we show a few results for different values of $p_1$. A careful study around the point $p_1=0.4888$ shows that we recover the critical character of the surviving probability at ($p_1=0.4888 \pm 0.0005, p_2=0$).
The critical exponent $\delta$ measured at the compensation point ($\delta = 0.25 \pm 0.01$) is different from that at DP. 

\begin{figure}
\includegraphics[width=0.5\textwidth ]{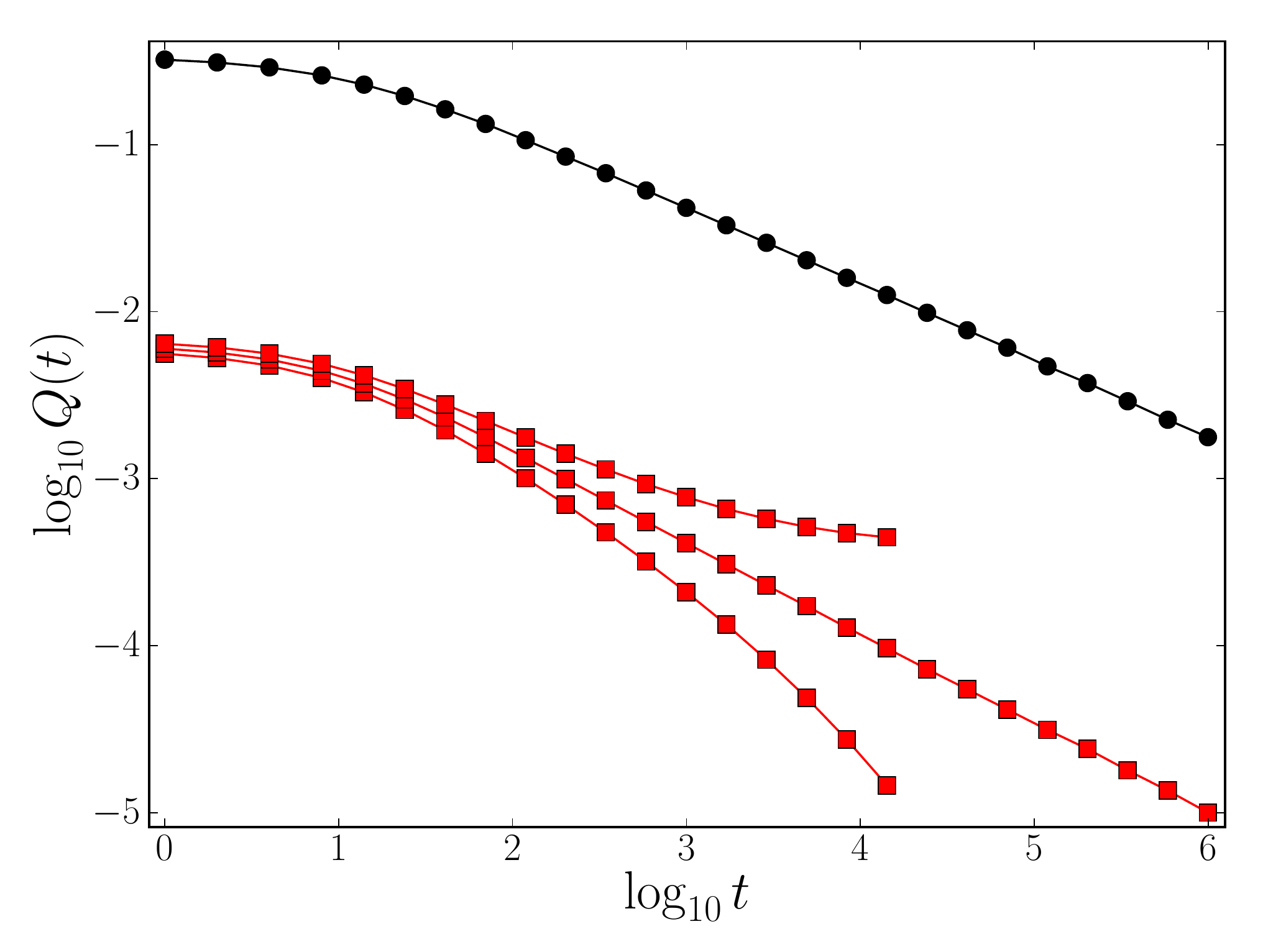}
\caption{(Color online). 
$Q(t)$ for different choices of $(p_1,p_2)$. 
Circles represent the pure DP at $p_1=p_2=p_{\text{subs}}=p_c$. 
Squares represent 
the AM with $p_1=0.01, p_{\text{subs}}=p_c$. From top to bottom, we used $p_2 =0.62, 0.60, 0.58 $. For $p_1=0.600$, $Q(t)$ displays a clear power-law with $\delta=0.53 \pm 0.01$. Averages are performed over about $10^8$ samples.
}
\label{qt2}
\end{figure}

The second compensation point is searched by setting $p_1=0.01$ and varying $p_2$.
In Fig. \ref{qt2}, we show the critical character of the point $(p_1=0.01, p_2=0.6000 \pm 0.0005)$. As for the previous point, this level of precision on the location of the critical point was obtained from a careful numerical study.
Similarly we find a new value for $\delta$: $0.53\pm 0.01$.

\begin{figure}
\includegraphics[width=0.5\textwidth]{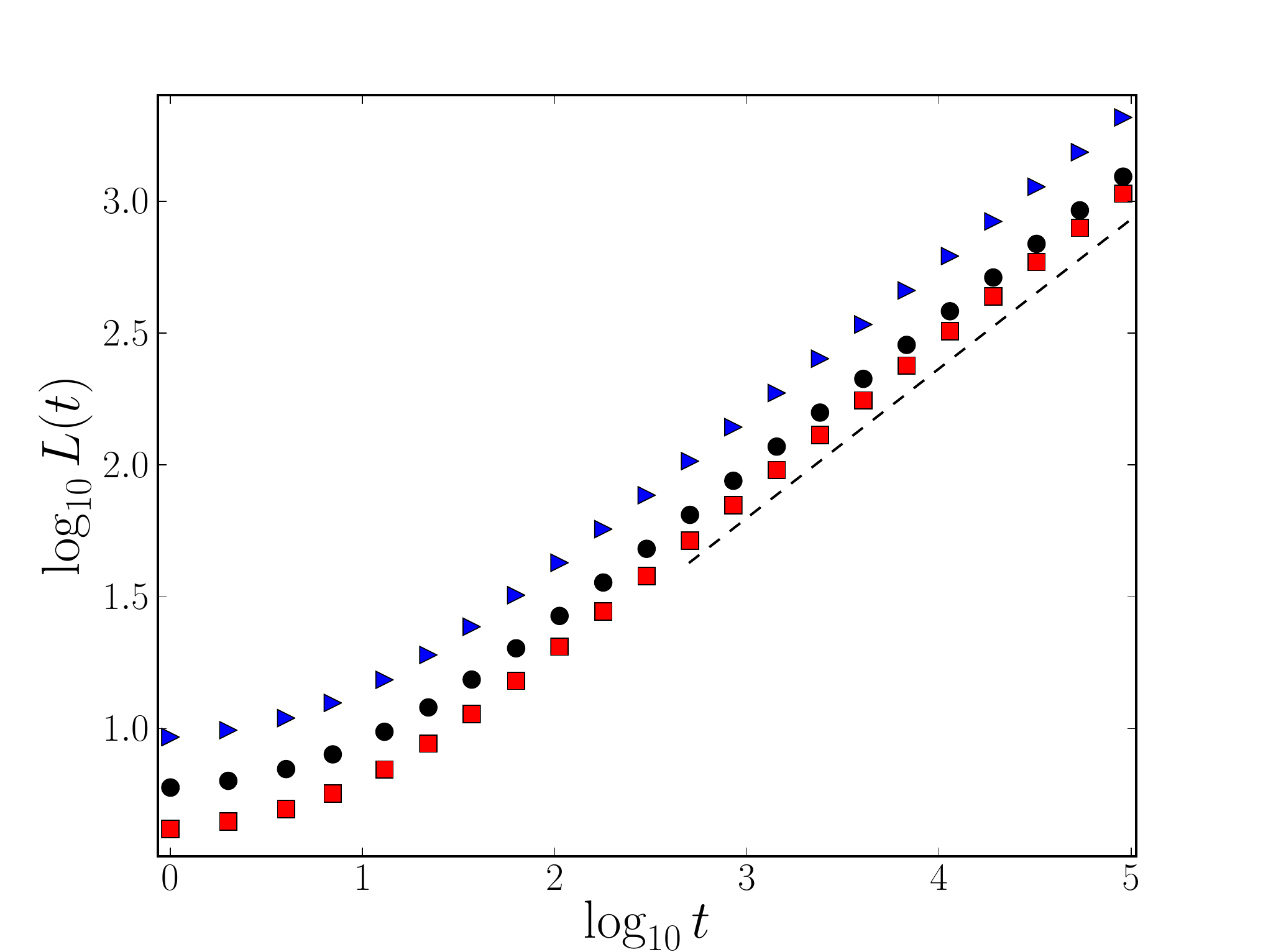}
\caption{(Color online). 
$L(t)$ for the AM at criticality. 
Circles represent the pure DP at $p_1=p_2=p_{\text{subs}}=p_c$. 
Squares represent the AM with $p_1=0.01, p_2=0.600, p_{\text{subs}}=p_c$. 
Triangles represent the AM with $p_1=0.4888, p_2=0, p_{\text{subs}}=p_c$.  
The dashed line corresponds to a slope $1/z=0.566$, using the exponent $z$ measured in pure  DP  (\ref{pureDP}). 
Averages are performed over $10^5-10^8$ samples. }
\label{Lt}
\end{figure}

\begin{figure}
\includegraphics[width=0.5\textwidth]{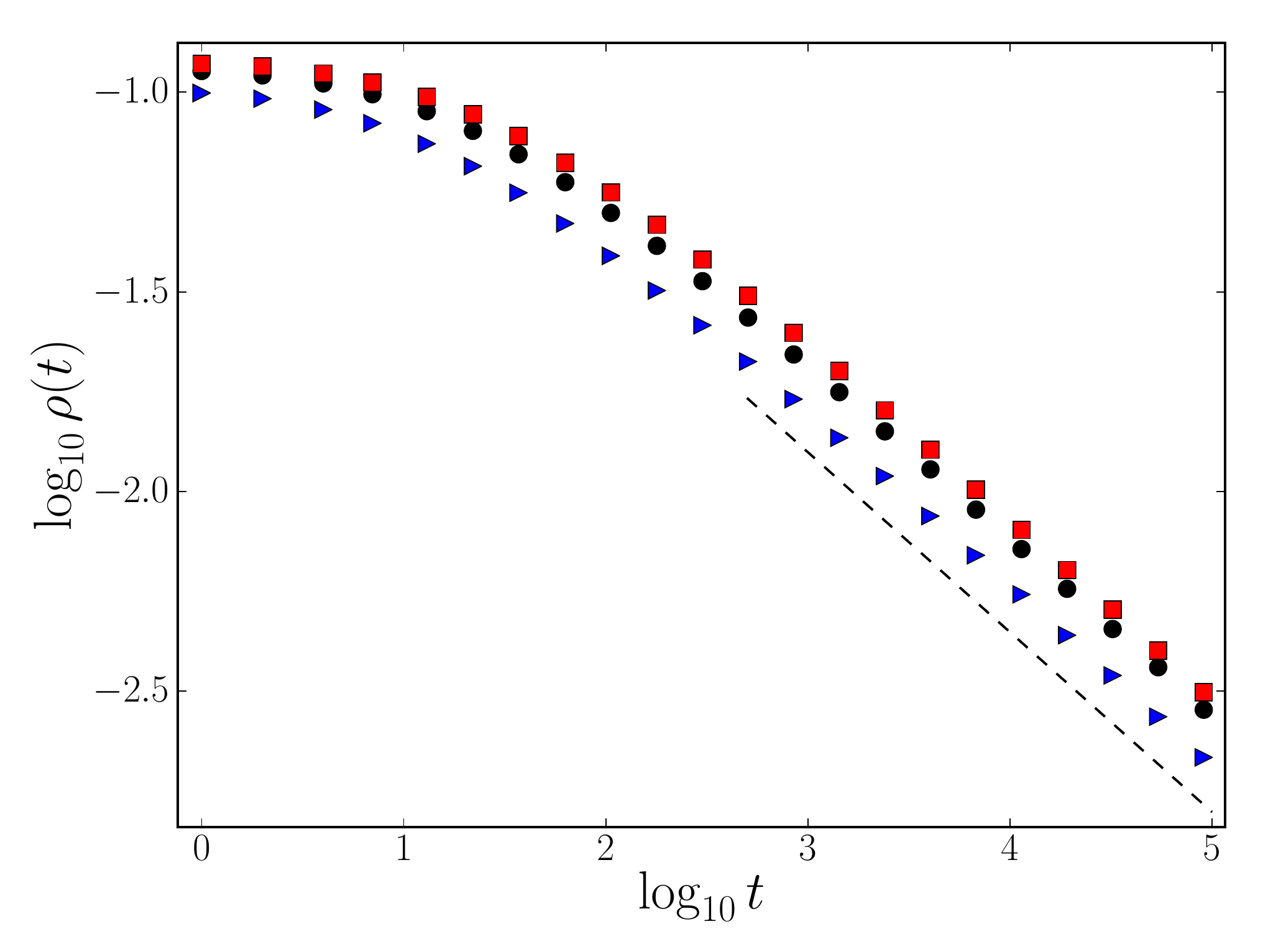}
\caption{(Color online). 
$\rho(t)$ for the AM at criticality. 
Circles represent the pure DP. 
Squares represent the AM with $p_1=0.01$, $p_2=0.600$, $p_{\text{subs}}=p_c$. 
Triangles represent the AM with $p_1=0.4888, p_2=0, p_{\text{subs}}=p_c$.  
The dashed line corresponds to the exponent  measured in pure  DP  (\ref{pureDP}).
Averages are performed over $10^5-10^8$ samples.}
\label{rhot}
\end{figure}

Let us present the critical behavior of the quantities related to the bulk exponents, $ \theta, z, \nu_\parallel$. 
$L(t)$ corresponds to the mean cluster width averaged over runs that survive until time $t$.
In Fig. \ref{Lt} we compare our data at two compensation points and at the  DP point:
we notice that the $z$ exponent does not change, unlike the coefficient before the power law.
In Fig. \ref{rhot}, $\rho(t)$ corresponds to the mean density averaged over runs that survived until $t$. 
The density  of a single run is measured as the ratio of the number of active sites at $t$ over the number of sites that were activated at least once.
Again, one may notice in Fig. \ref{rhot} that the exponent $\theta$ remains unchanged between the different critical points.

\begin{figure}
\includegraphics[width=0.5\textwidth]{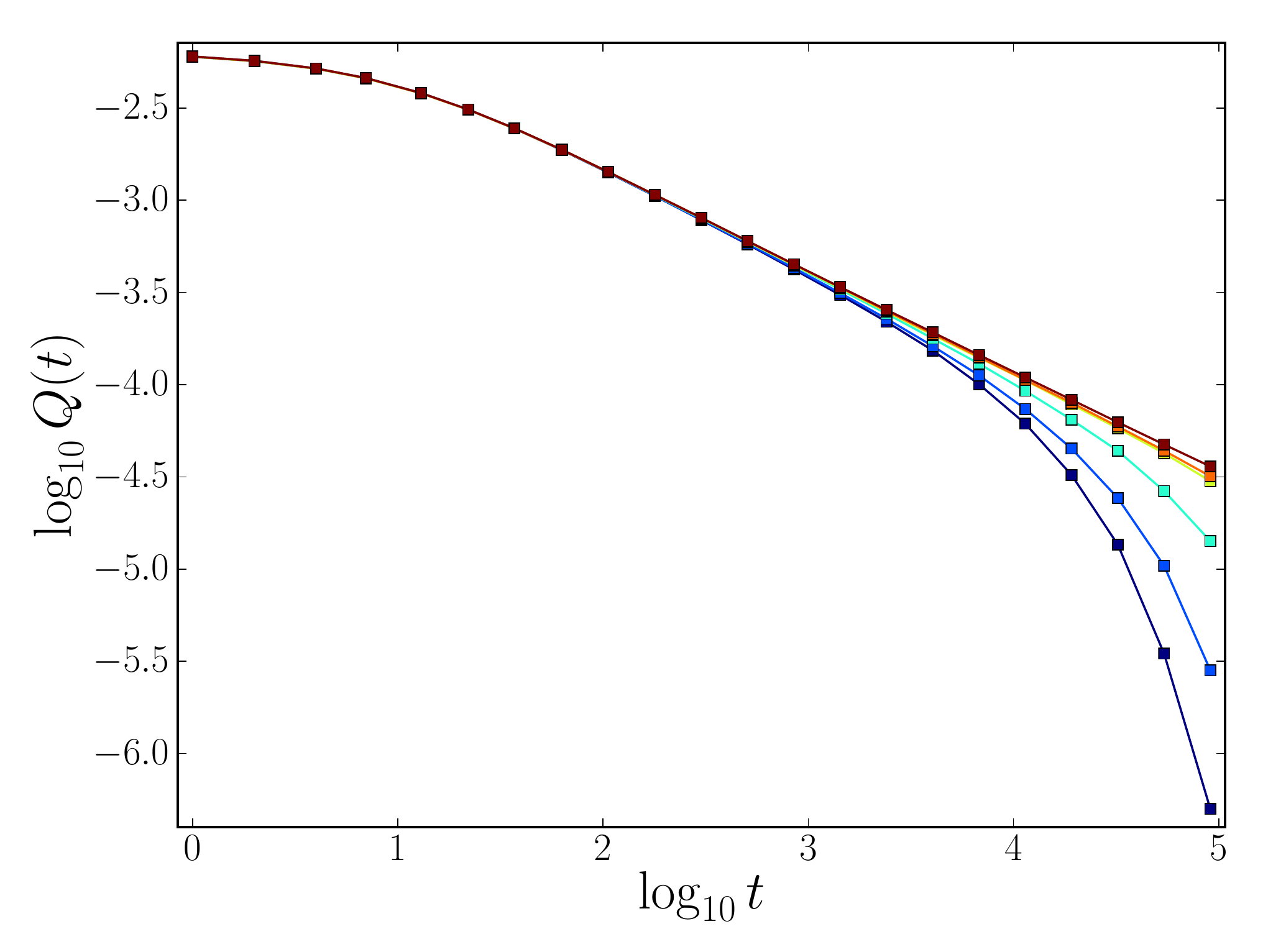}
\includegraphics[width=0.5\textwidth]{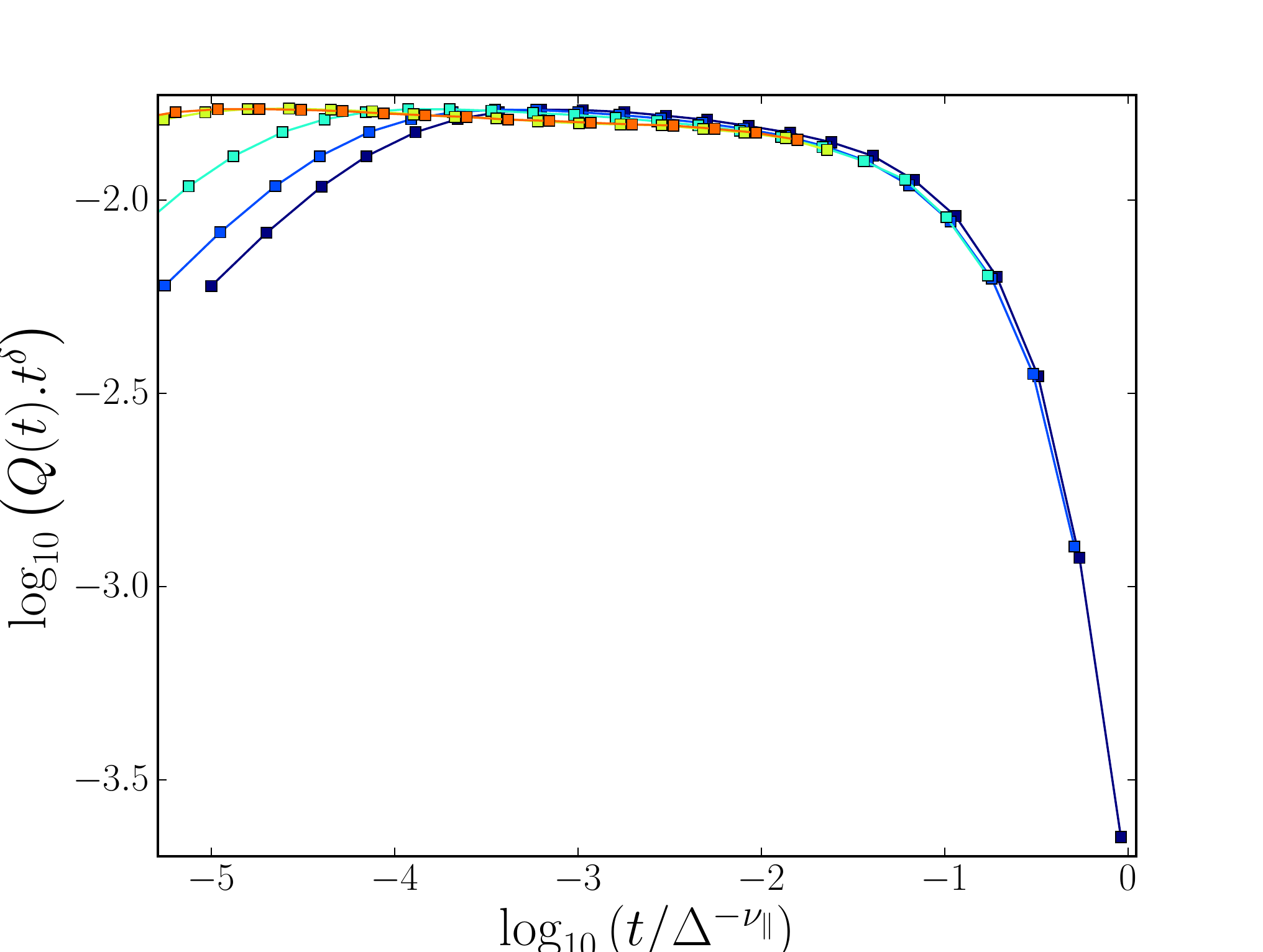}
\caption{ (Color online). 
Fig. (a): $Q(t)$ for the AM for $p_1=0.01$, $p_2=0.600$ and different $p_{\text{subs}}$. 
From top to bottom, $p_{\text{subs}}=0.287338$, $0.28733$, $0.28732$, $0.2873$, $0.2872$, $0.287$.
Averages are performed over $8\, 10^6$ samples. Fig. (b):
We collapse these data, plotting $Q(t) \cdot t^\delta$ against $t/ \Delta ^ {-\nu_\parallel}$.
We used the $\delta=0.53$ measured in figure \ref{qt2}, and the DP value given in (\ref{pureDP}) for $\nu_\parallel$.
}
\label{collapse}
\end{figure}

We want to check if $\nu_\parallel$ changes with $p_1$ and $p_2$.
To do this we set $(p_1=0.01, p_2=0.600)$ and use different values of $p_{\text{subs}} < p_c$ , thus varying $\Delta$, and observe the deviation from power-law behavior in Fig. \ref{collapse}(a). 
We consider the scaling law in Eq.(\ref{Q1}),
using the value of $\delta= 0.53$ extracted from Fig. \ref{qt2} and the DP value given in Eq.(\ref{pureDP}), 
we obtain a perfect collapse for the survival probability.
This shows that $\nu_\parallel$ does not change between compensation and DP. 

\begin{figure}
\includegraphics[width=0.5\textwidth]{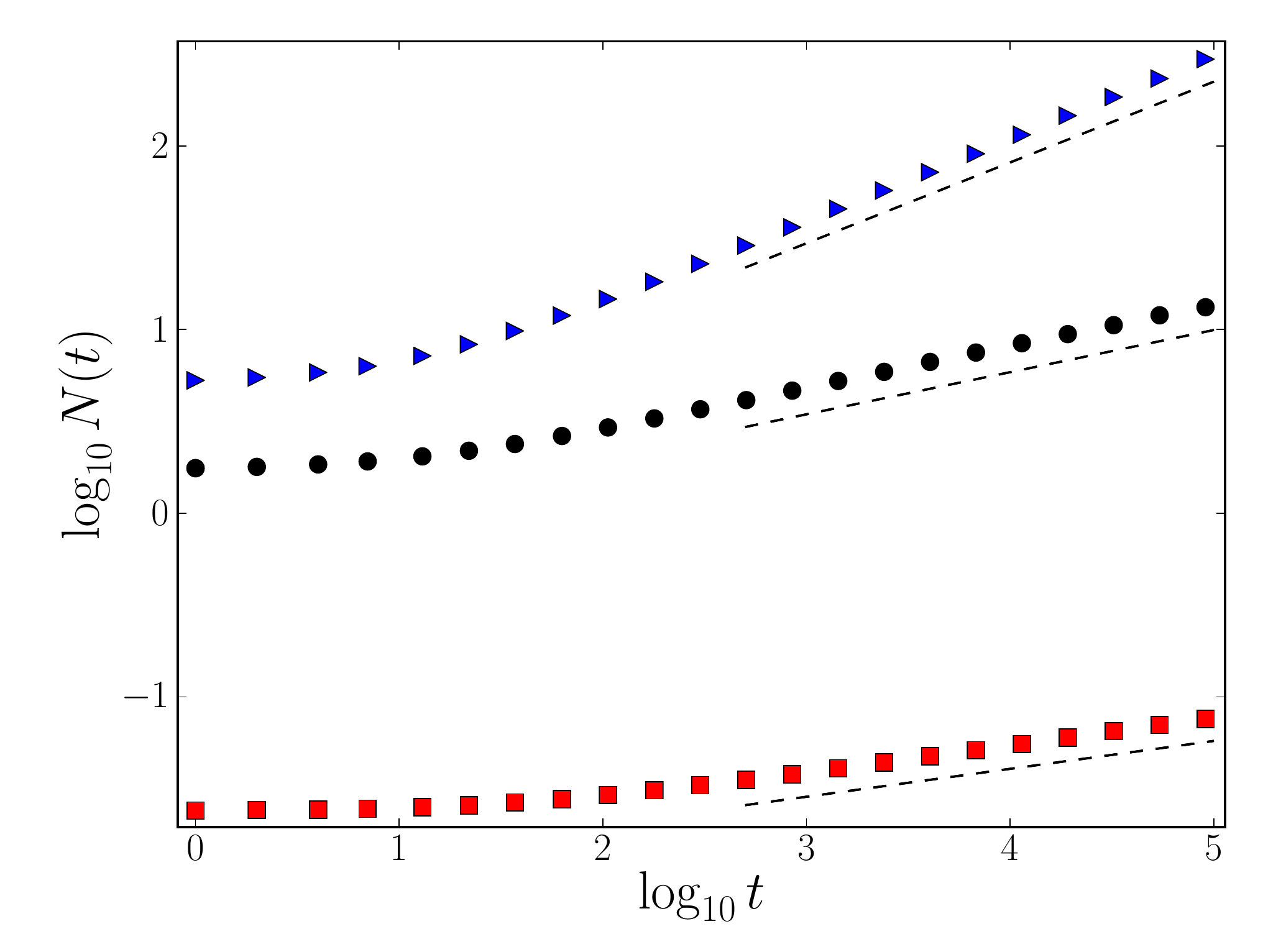}
\caption{(Color online). 
$N(t)$ for the AM at criticality. 
Triangles represent the AM with $p_1=0.4888, p_2=0, p_{\text{subs}}=p_c$.  
There we measure $\eta=0.44 \pm 0.01$ (dashed line). 
Circles represent the pure DP at $p_1=p_2=p_{\text{subs}}=p_c$. 
the dashed line corresponds  to the exponent  measured in pure  DP  (\ref{pureDP}).
Squares represent the AM with $p_1=0.01, p_2=0.600, p_{\text{subs}}=p_c$. 
We measure $\eta=0.15 \pm 0.01$ (dashed line).
Averages are performed over $10^5-10^8$ samples.
}
\label{Nt}
\end{figure}

The scenario is different for the spreading exponents $\delta, \eta$ and $ \tau$. 
We already saw that $\delta$ changes at compensation. In addition, in Fig. \ref{Nt},
the number of active sites averaged over all runs, $N(t)$, is seen to depend on the compensation pair $(p_1,p_2)$. For the compensated point $(p_1=0.4888, p_2=0)$ we measure $\eta= 0.44 \pm 0.01$ and for $(p_1=0.01, p_2=0.600)$ we measure $\eta= 0.15 \pm 0.01$.
At compensation, we expect the hyperscaling relation (\ref{hypers}) to hold. 
As $z$ and $\theta$ are found to be constant, the only way to preserve this relation is to have $\delta + \eta = d/z -\theta = \text{const}$. This constant is $0.680 \pm 0.002$, if we refer to \cite{Dickman1999}. 
For the point $(p_1=0.4888, p_2=0, p_{\text{subs}}=p_c)$, we find that $\delta+\eta = 0.69 \pm 0.02$.
For the other compensation point $(p_1=0.01, p_2=0.600, p_{\text{subs}}=p_c)$, we find $\delta+\eta = 0.68 \pm 0.02$.
These results are consistent with the expected value, for both compensation points. 

\begin{figure}
\includegraphics[width=0.5\textwidth]{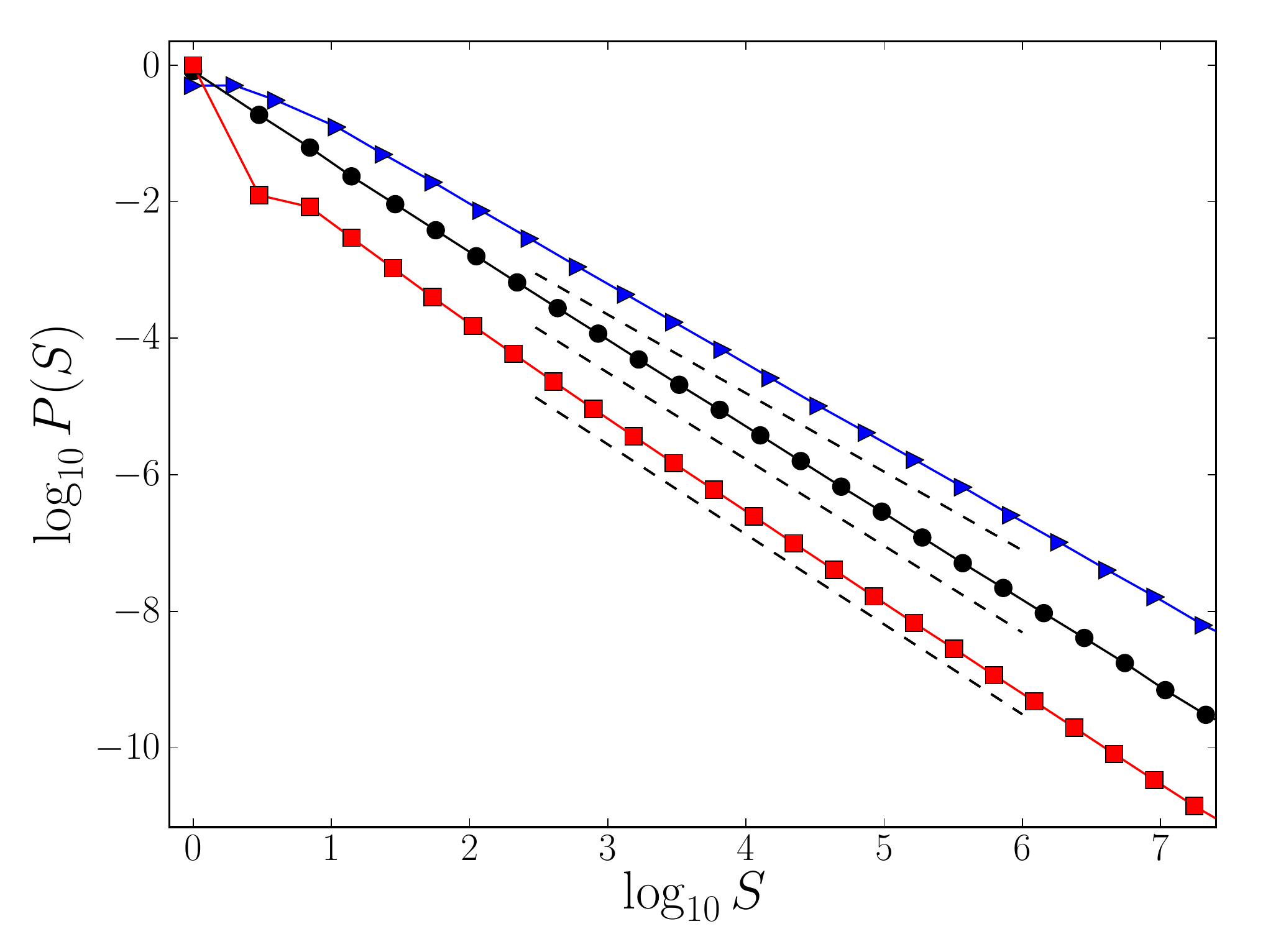}
\caption{ (Color online). 
$P(S)$ for the AM at criticality. 
Circles represent the pure DP at $p_1=p_2=p_{\text{subs}}=p_c$. 
We check that $\tau=1.268 \pm 0.005$.
Squares represent the AM with $p_1=0.01$, $p_2=0.600$, $p_{\text{subs}}=p_c$. 
We measure $\tau= 1.318 \pm 0.005$.
Triangles represent the AM with $p_1=0.4888$, $p_2=0$, $p_{\text{subs}}=p_c$.  
There we measure $\tau=1.151 \pm 0.005$. 
Averages are performed over $10^5-10^8$ samples.
}
\label{P(S)}
\end{figure}

In Fig. \ref{P(S)}, we present the probability density function $P(S)$.
The scaling relation (\ref{tau})  holds for the compensation process. 
In particular for the first compensation point $(p_1=0.4888, p_2=0)$, using $\delta= 0.25 \pm 0.01$, the equation (\ref{tau}), and $\delta+\eta = 0.69 \pm 0.02$, we expect $\tau=1.148 \pm 0.006$. We measure $\tau = 1.151 \pm 0.005$.
For the other compensation point we expect $\tau = 1.315 \pm 0.006 $ and measure $\tau = 1.318 \pm 0.005$. 
These results are all consistent with the  expected values, within our numerical precision.
We see that the relations derived in the first section are still valid, except for the time-reversal symmetry which is violated, since $\delta \neq \theta$.

\section{Discussion and Related Models}\label{discussion}

\begin{figure}
\includegraphics[width=0.4\textwidth]{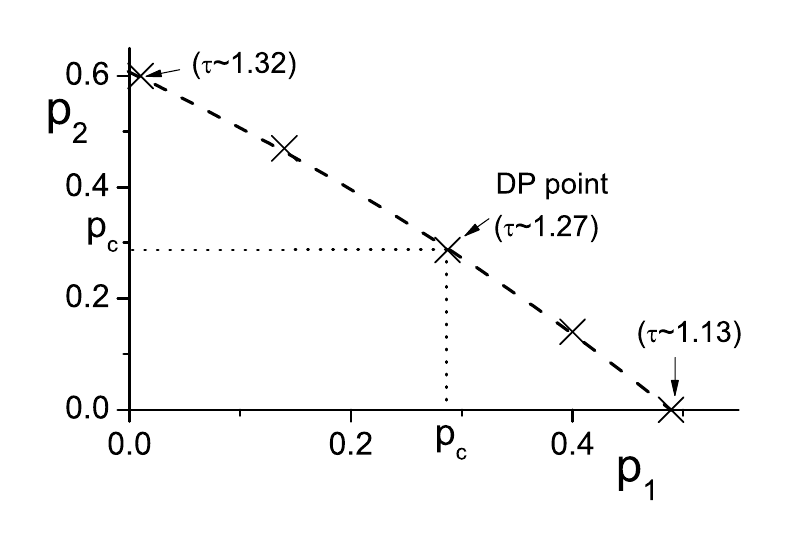}
\caption{
Phase diagram of the system in the $(p_1, p_2)$ parameter space with $p_{\text{subs}}=p_c=0.287338$.
The dashed line (schematic) is a critical line on which quantities in the system are power-law distributed. 
Above the line there is annular growth, and below there is sub-critical growth.
The bulk exponents $\theta, z, \nu_\parallel$ are equal to DP values all along this line, 
whereas the spreading exponents $\delta, \eta, \tau$ vary continuously along the line (representative values of $\tau$ are indicated).
The crosses correspond to those points along the critical line that were numerically determined.
}
\label{p1p2}
\end{figure}

A generic description of the behavior of the model can be presented in the  $(p_1, p_2)$  parameter plane (Fig. \ref{p1p2}).
In this plane there is a line along which the behavior of the system is critical.
This line passes through the DP point $(p_1=p_c, p_2=p_c)$.
The values of the bulk exponents  $z$, $\theta$ and $\nu_\parallel$ are constant all along the line.
The three spreading exponents $\delta$, $\eta$ and $\tau$ change continuously when we move along the line, but always respect the relations (\ref{hypers}) and (\ref{tau}), 
so that there is only one independent exponent that changes.
The value of $\delta$ passes from lower-than-DP values when $p_1>p_c>p_2$, to larger-than-DP values
when $p_1<p_c<p_2$.
Out of this line, there is in general a stretched exponential contribution to the distribution of the relevant quantities of the problem.

Although we do not have an analytical proof of our main claim, i.e. the existence of a critical line in the $(p_1, p_2)$ plane, we can simply demonstrate that there is a singular line in some respect. Along the diagonal of the $(p_1, p_2)$ plane, the DP point separates a long term survival probability $Q_{st}$ of zero (towards the origin, $p_1=p_2=0$) and a finite value of $Q_{st}$ (towards larger values of $p_1$ and $p_2$).
The values of $Q_{st}$ in other parts of the $(p_1, p_2)$ plane must smoothly match this known behavior.
In particular, we will have a singular line separating a region with $Q_{st}=0$, towards the origin and along this line, from another region with $Q_{st}\neq 0$, to the right and above this line. This proves that there is a singular line with respect to $Q_{st}$ in the $(p_1,p_2)$ plane.
Our expectation is that this singular line is also a critical line in which quantities are power law distributed.

\begin{figure}
\includegraphics[width=0.5\textwidth]{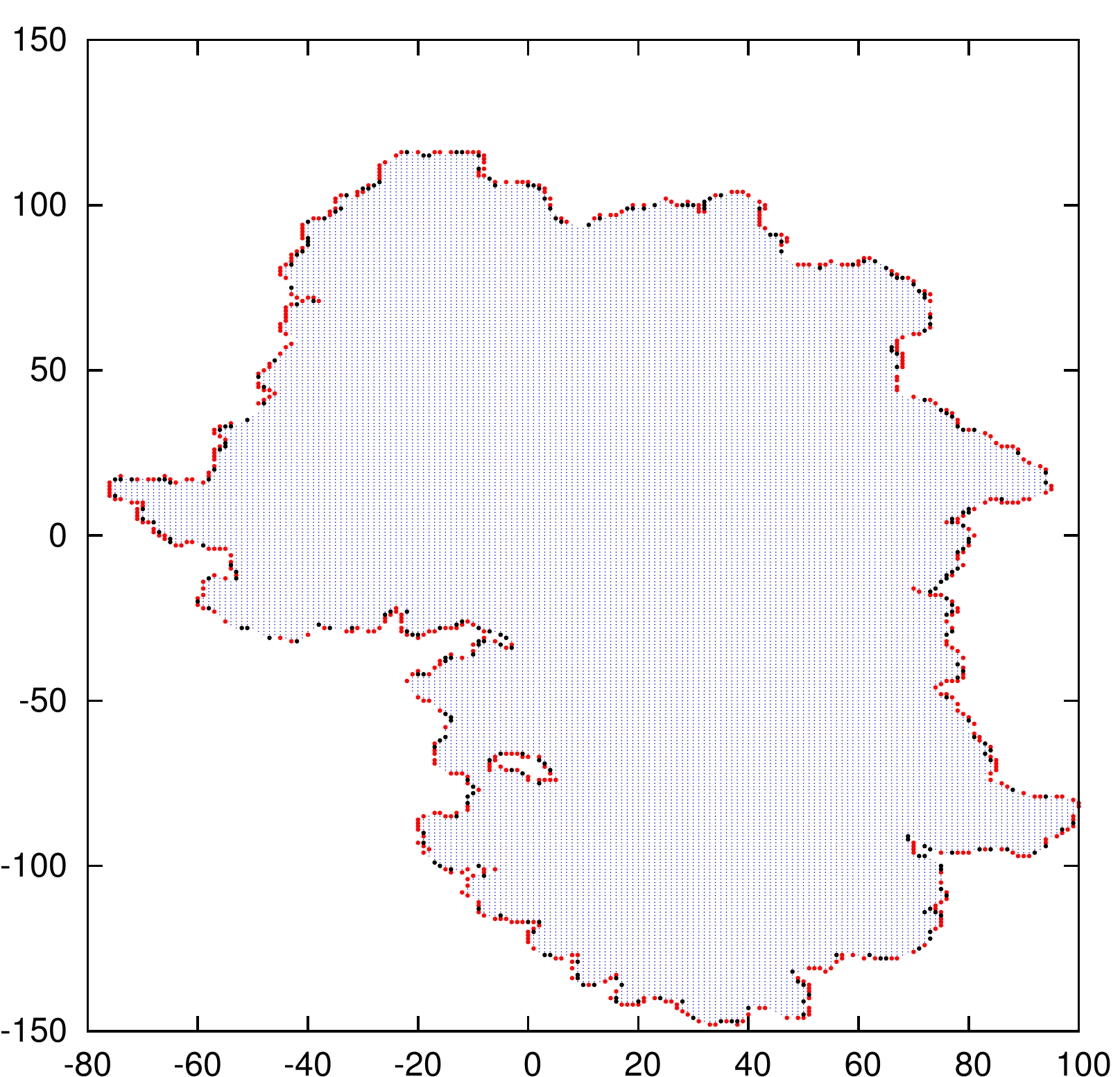}
\caption{(Color online). 
A snapshot of a growing cluster in the AM at compensation ($p_1=0.01$, $p_2=0.600$). Red (gray) and black points at the border are those sites that have been attempted  once or twice respectively. The whole interior is formed by sites that have been attempted more than twice.}
\label{cluster}
\end{figure}
We can understand the behavior of the bulk exponents if we think that these exponents can 
be measured starting from a fully active lattice.
In this case the evolution of the system coincides with that of pure DP after a few time steps. 
However, bulk exponents can also be measured on the surviving runs started from a single active site.
In this case space-time is divided in two regions: the active one, and the outer, inactive one. 
In Fig. \ref{cluster} we show a snapshot of a AM growing cluster at a given $t$. 
We see that sites that make the difference with usual DP  are mostly located at the boundary of the active region.
We consider a large box  of size $\ell_\perp \ll \xi_\perp$ in space and $\ell_\parallel \ll \xi_\parallel$ in time, sufficiently far away from the boundary with  the inactive region. 
Its statistical properties will be completely independent of its precise location and 
are indistinguishable from those of a box with the same size, with the fully active initial condition.
Since the role of the boundaries is asymptotically small, this shows that the bulk exponents $\theta, \nu_\perp, \nu_\parallel$ and thus $z$ are unchanged by the compensation process also if we use the single seed  initial condition.
However, the spreading exponents $\delta, \eta, \tau$ 
are naturally defined only in the seed initial condition, and involve averages over all runs.  These exponents depend continuously on $p_1$ and $p_2$.

A similar scenario happens in $1$-dimensional models which display critical behavior, despite their breaking of the time-reversal symmetry. 
In these models \cite{Mendes1994a, Jensen1993a, Jensen1993, Marques1999, Dickman1999, Odor1998, Park2007, Munoz1996, Munoz1998} each site is active or inactive, as in DP, but is equipped with an additional auxiliary field $\phi$ which allows for a large degeneracy of the absorbing state.
We discuss  DP with auxiliary fields using the example of the Threshold Transfer Process (TTP) \cite{Mendes1994a, Odor1998}.
In the $1$-dimensional TTP, a site may be vacant, singly or doubly occupied, 
corresponding respectively to states $\sigma_i = 0, 1$ or $2$.
The auxiliary field $\phi$ denotes the density of singly occupied sites. 
A doubly occupied site corresponds to the active state. 
Initially, only the site at the origin is doubly occupied,
while the state $\sigma_i$ of each other site is $1$ with probability $\phi_{\text{init}}$, and $0$ otherwise. 
At each time step, a site $i$ is selected at random.
If $\sigma_i(t) < 2$, then $\sigma_i(t+1) = 1$ with probability $r$ and $0$ with probability $1-r$, irrespective of the precise initial value. 
If $\sigma_i(t) = 2$, the site releases one particle to all neighbors with $\sigma(t) < 2$. 
Contrary to the DP case, there are infinitely many absorbing states since any configuration with no doubly occupied site is absorbing.

In TTP, $r$ plays the role of control parameter, and in $d=1$, $r_c = 0.6894$ \cite{Odor1998}. At criticality the bulk exponents and the hyperscaling relation behave as in DP, independently of the initial condition. 
However the spreading exponents continuously depend on the initial condition $\phi_{\text{init}}$.
Setting the initial density of singly-occupied sites to its stationary value $\phi_{st}=r_c$, one recovers the full set of DP critical exponents \cite{Mendes1994a}.
As far as we know, a theoretical explanation for the continuous change in the spreading exponents $\delta, \eta$ is still an open question.

It is worth mentioning a second class of models with similar behavior, which corresponds to DP with special absorbing boundary conditions.
In particular DP with  absorbing walls at positions $x(t) = \pm C \cdot t^{1/z}$ shows spreading exponents that continuously depend on $C$ \cite{Kaiser1995, Kaiser1994}. Analogous results with a moving active wall are presented in \cite{Chen1999}.
Moreover, one dimensional models with soft or hard walls conditions can be studied analytically  in the case of Compact DP.
They can be mapped onto compact first attempt (for soft walls) and compact first infection (for hard walls). 
Dickman  showed \cite{Dickman2001a} that in this case the critical behavior is maintained when $p_1$ is reduced, i.e. in this case we do not have a stretched exponential contribution.

In conclusion, memory effects in immunization problems, or the presence of auxiliary fields in TTP-like models, introduce high degeneracy of the absorbing state and thus break the time reversal symmetry. 
In these systems, at criticality, the bulk DP exponents are recovered. However, if the initial condition is sufficiently far from its stationary value (which is $\phi_{\text{st}}$ for TTP-like models, and the fully twice-attempted lattice for the compensation model, or the fully once-infected lattice in the modified first Infection Model),
the spreading exponents depend continuously on the initial condition. 
Non-stationarity seems to play a key role in the observed anomaly of the spreading exponents.

\section{Conclusions}\label{conclusion}

We have shown that DP universal behavior is strongly affected by changes in the first probabilities to activate sites. 
This modification corresponds to a special case of ``long memory", where each site remembers 
exactly how many times it has been activated (or attempted) before. Our main result is that, although the change of the very first attempt probability takes the model out of criticality, by changing the second attempt probability in the opposite direction, we can restore critical behavior, in a process we called compensation.

We have focused in this paper on the case of two spatial dimensions, but
qualitatively the same behavior is obtained for one dimension. However in one dimension the deviations from criticality when $p_1$ is changed are much weaker than in two dimensions, making the determination of the compensation condition much more difficult numerically.

It is uncertain for us at present if the phenomenology of the compensation in the Attempt Model applies also to the Infection Model. It seems that the compensation effect can occur, but we do not have enough numerical evidence to assure that the systematics of critical exponents is the same as that discussed in this paper for the Attempt Model.

The quantitative change we have obtained of the critical exponent $\tau$ is rather weak. 
For instance we have obtained changes of the value of $\tau$ of approximately $\pm 0.1$ around the DP value $\tau=1.268$. These
variations are typically within the error bars of experimentally determined exponents in concrete situations. 
Although changes in other exponents like $\eta$ and $\delta$ were found to be larger,
in some practical situations where avalanches are observed, only the size distribution exponent $\tau$ can be directly measured. 
So it seems dubious that the effect we have discussed can be observed in a concrete realization. 

In this respect we want to mention that we have found other realizations of the DP process 
where the effect is quantitatively much more important. 
For instance, the process in which we try to activate neighbors with probability $p$, 
having in addition a self-activation probability $p_0$ of the same site, belongs also to the DP universality class. 
In this case we have observed that a lower probability to activate neighbors for the first time can be compensated by larger self-activation probabilities during the next steps,
and in this case the quantitative effect is much more important. 
In particular we have obtained avalanche size distributions with $\tau$ as large as $\simeq 1.7$. This variant of the DP problem and its relation to earthquake dynamics will be addressed in another presentation.

Aside from any application to a concrete situation, we want to stress the fact that the present model provides a link between two classes of models with very different behavior: the models with auxiliary field and the Infection Model. Although we obtained the same results as in models with an auxiliary field (criticality  with  scaling relations preserved, time-reversal symmetry broken), our microscopical description fits in the framework of modified First Infection Models, for which analytical computations have been successful\cite{Jimenez-Dalmaroni2003}. This may be an interesting approach to the open problem of initial-condition-dependent exponents in absorbing phase transitions.

\section{Acknowledgments}
EAJ is financially supported by Consejo Nacional de Investigaciones Cient\'{\i}ficas y T\'ecnicas (CONICET), Argentina. Partial support from grant PIP/112-2009-0100051 (CONICET, Argentina) and by ANR grant 09-
BLAN-0097-02. is also acknowledged.


\begin{thebibliography}{10}
\bibitem{Footnote1}
More precisely the field theory of absorbing phase transitions shows \cite{Grassberger1983c} that the density exponent 
$\beta$ is associated with the annihilation operator while the survival exponent $\beta^ \prime$ is associated with the creation operator.
\end{thebibliography}
\end{document}